# Optimizing Geant4 Hadronic Models

*Krzysztof* Genser[1], *Soon Yung* Jun[1], *Alberto* Ribon[2], *Vladimir* Uzhinsky[3], and *Julia* Yarba[1,*]

[1]Fermi National Accelerator Laboratory†, Batavia, IL, 60510-5011, USA
[2]CERN 27210, CH-1211 Geneva, Switzerland
[3]Joint Institute for Nuclear Research, Dubna, Moscow Region, Russia 141 980

**Abstract.** Geant4, the leading detector simulation toolkit used in high energy physics, employs a set of physics models to simulate interactions of particles with matter across a wide range of energies. These models, especially the hadronic ones, rely largely on directly measured cross-sections and inclusive characteristics, and use physically motivated parameters. However, they generally aim to cover a broad range of possible simulation tasks and may not always be optimized for a particular process or a given material. The Geant4 collaboration recently made many parameters of the models accessible via a configuration interface. This opens a possibility to fit simulated distributions to the thin target experimental datasets and extract optimal values of the model parameters and the associated uncertainties. Such efforts are currently undertaken by the Geant4 collaboration with the goal of offering alternative sets of model parameters, also known as "tunes", for certain applications. The effort should subsequently lead to more accurate estimates of the systematic errors in physics measurements given the detector simulation role in performing the physics measurements. Results of the study are presented to illustrate how Geant4 model parameters can be optimized through applying fitting techniques, to improve the agreement between the Geant4 and the experimental data.

## 1 Introduction

Geant4 [1–3] is a toolkit for simulating the passage of particles through matter that is widely used in high energy physics as well as a number of other domains. In high energy physics it is used to design detectors and to optimize calibration and reconstruction software, and to simulate Monte Carlo physics events for certain experimental setups. In domains such as neutrino physics it is also used to simulate the neutrino beamlines and to predict the neutrino flux through detectors.

Geant4 offers a set of physics models to simulate interactions of particles with matter across a wide range of interaction energies. These models, especially the hadronic ones, rely largely on directly measured cross sections and phenomenological predictions with physically motivated parameters estimated by theoretical calculations or measurements. However, in general they aim at covering a wide range of simulation tasks and are not always optimized for a given process or a given material.

*e-mail: yarba_j@fnal.gov

This brings up important questions :

- How sensitive are the Geant4 predictions to variations in model parameters ?
- How will it translate into the simulation of a detector design, and subsequently into the experimental observables ?

Starting with release 10.4, the Geant4 collaboration has made it possible to access at run time some of the underlying parameters in several models, and to vary these parameters. This opens a possibility to explore how the Geant4 outcome changes with variations of model parameters which can be illustrated by figure 1. Subsequently it paves way to fitting the simulated distributions to experimental datasets and extracting optimal values of the parameters and the associated uncertainties.

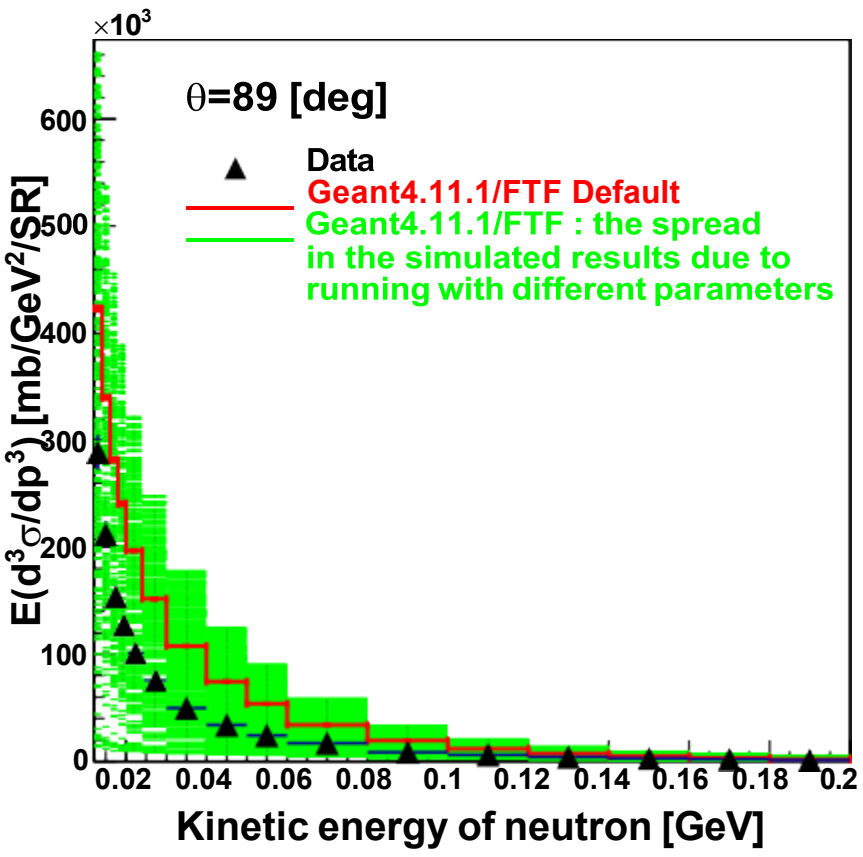

**Figure 1.** Lorentz invariant cross section as a function of the kinetic energy of secondary neutrons produced in proton-Lead interactions at 7.5 GeV/c, at the polar angle of the secondary of 89 degree. The red curve shows the default Geant4.11.1 results as simulated by the Fritiof [4] model. The green area represents the spread in the simulated results due to running with different settings of the Fritiof model parameters. The black triangles represent the data by the ITEP771 [5] experiment.

## 2 Initial Phase of Exploring Geant4 Model Parameters

We started with the initial set of configurable parameters that was available as of release 10.4, for such key hadronic models as Fritiof, Bertini-like cascade [6], and PreCompound [7]. A detailed description of the study can be found in [8]. In short, it can be summarized as follows:

- Varying and optimizing parameters of the Geant4 models generally leads to better agreement with some experimental data.
- However, the number of configurable parameters available at the time, per model, were too few to reach a better agreement across the board.

Subsequently the Geant4 collaboration has decided to focus on one model at a time, expands the model's configuration interface, and explore in greater details how the parameters of each model can be further optimized through fits to experimental data. Currently we focus on the Fritiof model as described in Sect. 3.

## 3 Studying Parameters of the Geant4 Fritiof Model

Being the recommended hadronic model of Geant4 to use at high energies, the Fritiof model (FTF) is based on modeling non-diffractive or diffractive interactions, quark exchange processes, and quark-gluon string formation, with subsequent string fragmentation through the LUND mechanism. Its validity range spans from 3 GeV and into the TeV range. The model currently remains in active development, with many features being refined and new features being added. In addition to [4], its thorough description, including its configuration interface and many technical implementation details, can be found in [9] and [10].

Our current study focuses on modeling such processes as nuclear target destruction and quark exchange without or with excitation of participants. Later in this paper we will show how parameters involved in modeling of these processes can be optimized through fits to the experimental data.

## 4 Experimental Datasets

There is a large volume of experimental data that can be used to tune hadronic models of Geant4. So far we have chosen a number of thin target datasets from such experiments as ITEP771 [5], Ishibashi *et al.* [11], HARP [12], and NA61/SHINE [13, 14].

The collection of experimental data that are currently included in the study covers a range of energies from 3 GeV and up to 60 GeV for such beam particles as proton or charged pions interacting with various types of nuclear targets. The data has been accumulated by various experiments over the past several decades. The datasets are summarized in table 1.

**Table 1.** Datasets included in the Geant4 model parameter fits.

| Experiment | Projectile | Target | Final State | Observable |
|---|---|---|---|---|
| ITEP771 | 7.5 GeV/c proton, 5 GeV/c $\pi^\pm$ | C, Cu, Pb | pX, nX | $Ed^3\sigma/d^3p$ |
| K.Ishibashi et al. | 3 GeV proton | C, Fe, Pb | nX | $d^2\sigma\ d\theta dE$ |
| HARP | 3-12 GeV/c proton or $\pi^\pm$ | C, Cu, Pb | $\pi^\pm$X, pX | $d^2\sigma/d\theta dp$ |
| NA61/SHINE | 31 GeV/c proton, 60 GeV/c $\pi^+$ | C | $\pi^\pm$X | $d^2\sigma/d\theta dp$ |

## 5 Fitting Package

The fitting package of our choice has been the Professor Tuning Toolkit [15]. In short, the idea of Professor is the following:
A set of parameter values represents a point $P_i$ in the multi-parameter space; there may be many such points. One can randomly sample through such points, within physically meaningful range of values. For every such point one can simulate the data combinatorics that are defined by how many experimental datasets one wants to include in the study, i.e., the beam types, the beam energies, with what target the beam is interacting, etc. Then for each $P_i$ one analyzes the simulated statistics, and the resulting histograms can be benchmarked against the experimental distributions. Each given bin of an observable distribution, as it varies from one set of parameters to another, is considered as a function of $P_i$ and is parametrized by a $n$-degree polynomial (the degree can be chosen by users but the default is the 3rd order). The resulting function is used to fit experimental data through traditional chi-square minimization. This approach substantially reduces the time required for producing dataset predictions.

## 6 Selected Results

We have performed "global fits" where several FTF parameters involved in modeling processes referred to in Sect. 3 were fit to multiple experimental datasets simultaneously. The fits were done separately for the proton or pion projectile given that the physics of the interactions is not the same for these beam types, thus the values of the model parameters should be different.

Figures 2 and 3 show results on neutron production in proton-nuclei interactions at several GeV/c of beam momenta, both simulated and experimental. To be more specific, in this domain, parameters involved in modeling the nuclear target destruction process in FTF have a significant impact. The benchmark against the experimental data clearly demonstrates how the use of best fit values of the model parameters improves the agreement between the Monte Carlo and the data significantly, as compared to the default Geant4.11.1/FTF simulation. Figure 4 shows results on $\pi^-$ production in proton-Carbon interactions at 31 GeV/c of beam momentum. Here parameters involved in modeling FTF quark exchange process have a significant impact on the simulated results. Again, the distributions simulated with best fit values of the model parameters are substantially closer to the experimental data, as compared to the default Geant4.11.1/FTF results. Results presented in Figures 2, 3, and 4 are all part of the same fit against multiple experimental dataset obtained for the proton projectile at different beam energy or momenta, interacting with different nuclear targets.

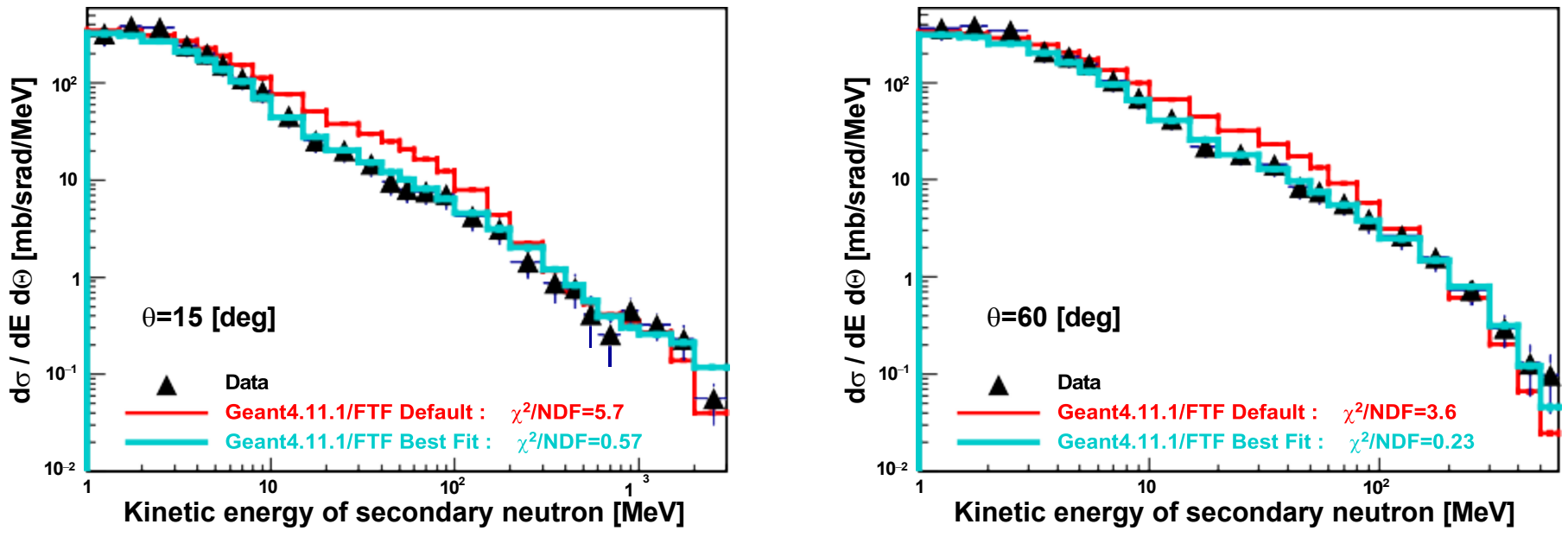

**Figure 2.** Production cross section as a function of the kinetic energy of secondary neutrons produced in proton-Lead interactions at 3 GeV of beam energy, at different values of the secondary's polar angle. The red curve represents results simulated with the default settings of the Geant4.11.1/FTF parameters. The cyan curve shows Geant4.11.1 results obtained with best fit values of the FTF parameters. The black triangles represent the data from Ishibashi et al. [11].

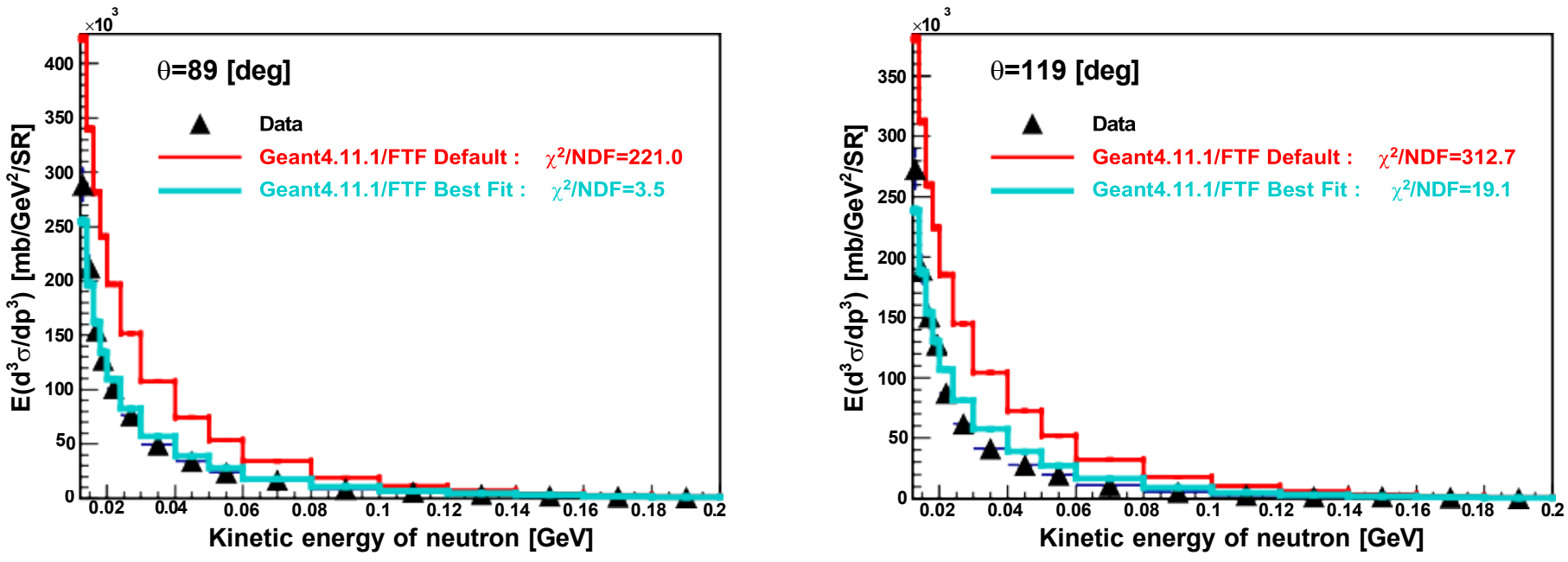

**Figure 3.** Lorentz invariant cross section as a function of the kinetic energy of secondary neutrons produced in proton-Lead interactions at 7.5 GeV/c of beam momentum, at different values of the secondary's polar angle. The red curve represents results simulated with the default settings of the Geant4.11.1/FTF parameters. The cyan curve shows Geant4.11.1 results obtained with best fit values of the FTF parameters. The black triangles represent the data by the ITEP771 [5] experiment.

Figure 5 shows results on pion production in $\pi^+$–Carbon interactions at 60 GeV/c of beam momentum. They are part of another fit performed against the collection of experimental data for the $\pi$ projectile of various energies or momenta, interacting with various nuclear targets. The distributions that are simulated with the best fit values of FTF model parameters show a much better agreement with the data, as compared to the default Geant4.11.1/FTF results.

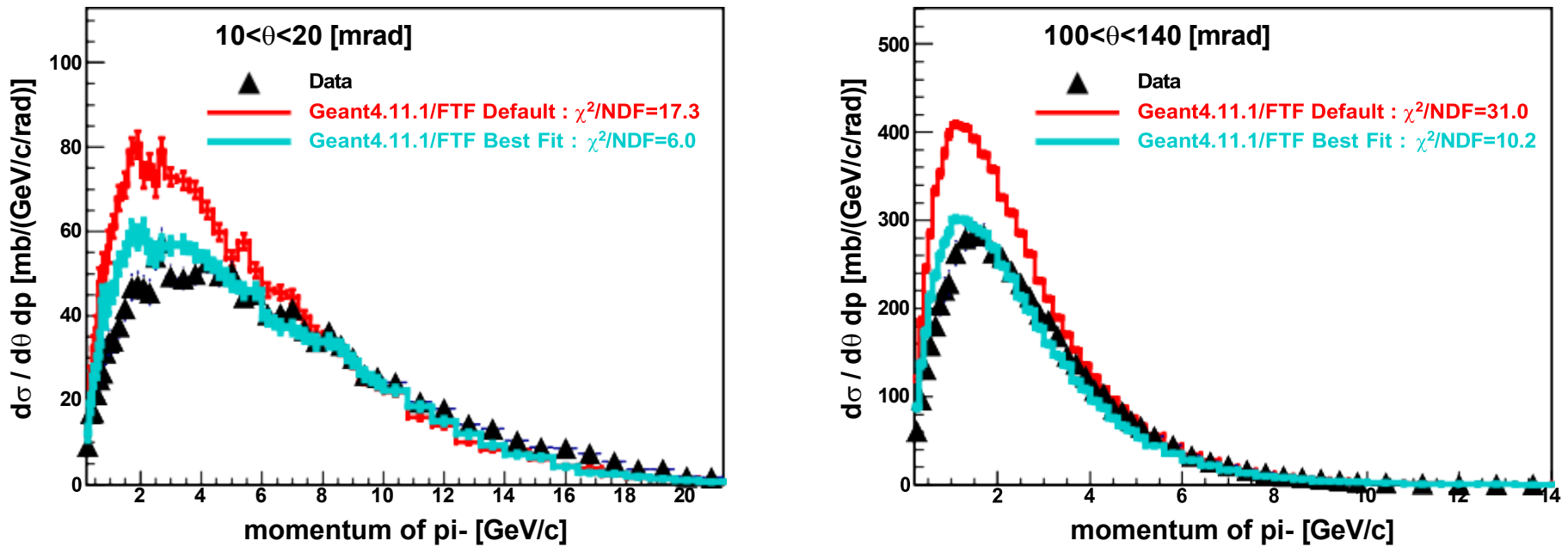

**Figure 4.** Production cross section as a function of the momentum of secondary $\pi^-$ produced in proton-Carbon interactions at 31 GeV/c of beam momentum, at different values of the secondary's polar angle. The red curve represents results simulated with the default settings of the Geant4.11.1/FTF parameters. The cyan curve shows Geant4.11.1 results obtained with best fit values of the FTF parameters. The black triangles represent the data by the NA61/SHINE [13] experiment.

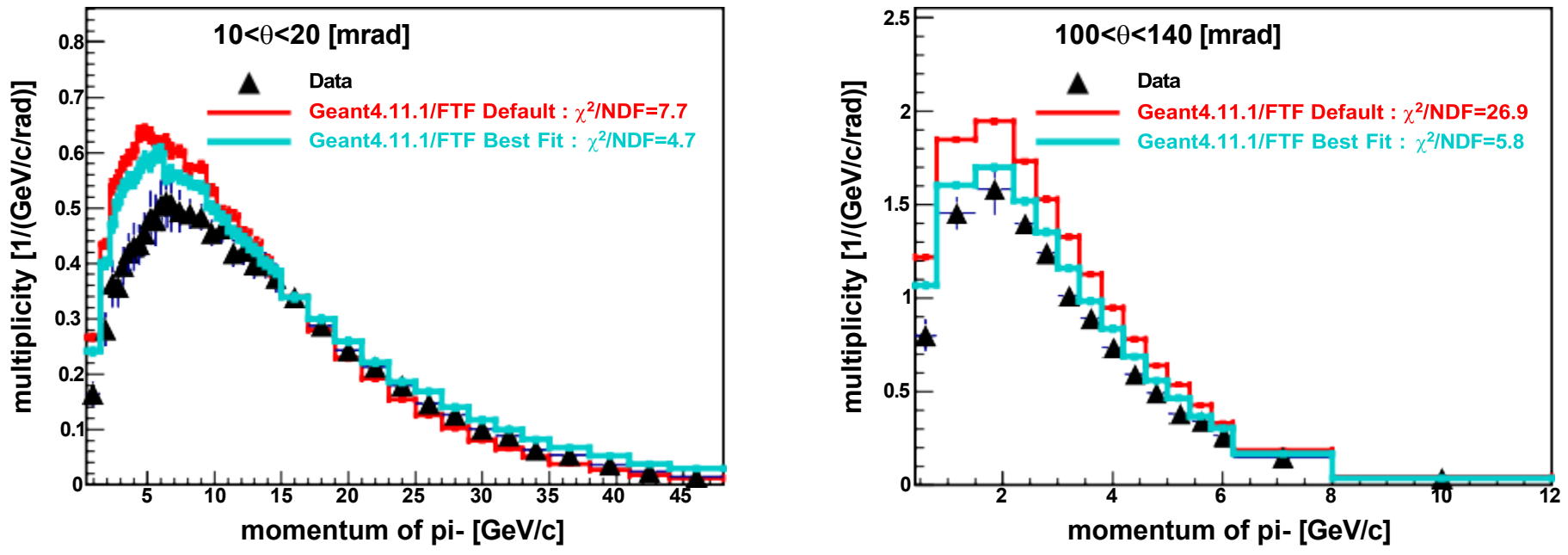

**Figure 5.** Multiplicity as a function of the momentum of secondary $\pi^-$ produced in $\pi^+$-Carbon interactions at 60 GeV/c of beam momentum, at different values of the secondary's polar angle. The red curve represents results simulated with the default settings of the Geant4.11.1/FTF parameters. The cyan curve shows Geant4.11.1 results obtained with best fit values of the FTF parameters. The black triangles represent the data by the NA61/SHINE [14] experiment.

## 7 Introducing Tunes in Geant4

Recently, along the course of the study, the Geant4 collaboration started considering the idea of developing the so called "tunes", for some specific study cases. Here a "tune" is a group of parameter settings collectively obtained through fits vs thin target data; they should be used as a group because some parameters are correlated. As an example, one can say that parameter settings that are suitable for simulating hadronic showers in the LHC calorimeters may be not quite optimal for modeling neutrino beamlines and neutrino fluxes, and vice versa.

In the most recent release Geant4.11.1 we have introduced some very preliminary FTF tunes for the baryon or pion projectile. We plan to explore whether FTF tunes can be added for other projectile particle types, and perhaps for different energy ranges. This work is in progress and more extensive validation is needed.

## 8 Future Plans

In the near future we plan to continue tuning the FTF model parameters, focusing on the areas where no significant progress has yet been made. An example of a such study case is illustrated by figure 6. It shows that the Geant4/FTF simulated distribution is very different even shapewise from the experimental one. Varying those FTF model parameters that we have so far been focusing on does not result in shifting the shape of the Monte Carlo distribution in the right direction. However, more extensive exploring of the FTF components and the associated parameters is likely to bring the desired results. Our future plans also include expanding the collection of experimental data involved in the study, e.g. more results from such experiments as NA61/SHINE [13, 14] at CERN or EMPHATIC [16] at Fermilab. Additionally, we plan to extend the work towards other Geant4 hadronic models. We also plan to explore the new tuning packages that become available.

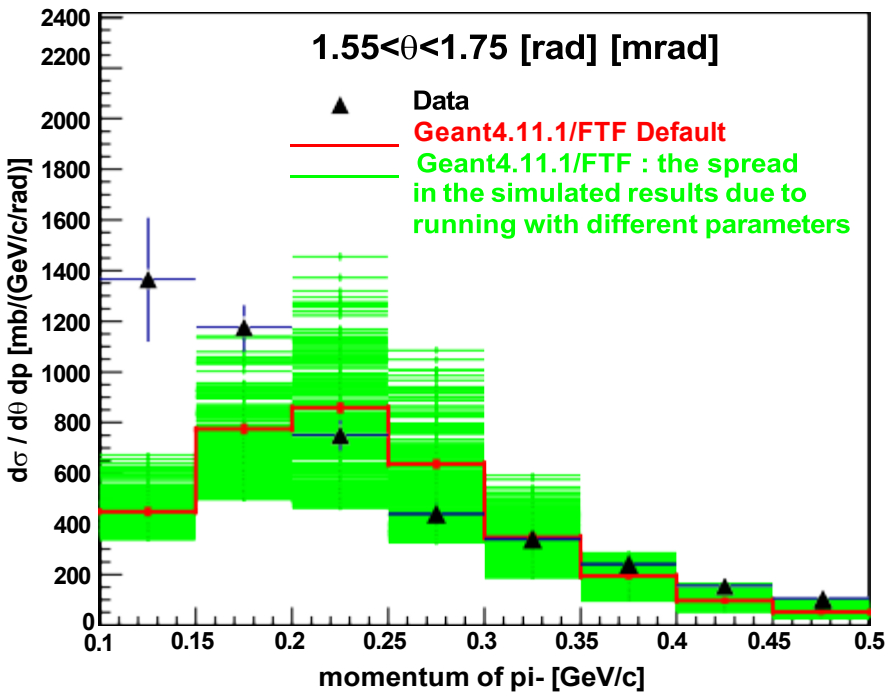

**Figure 6.** Production cross section as a function of the momentum of secondary $\pi^-$ produced in proton-Lead interactions at 5 GeV/c, at the polar angle of the secondary between 1.55 and 1.75 rad. The red curve shows the default Geant4.11.1/FTF results. The green area represents the spread in the simulated results due to running with different settings of the Fritiof model parameters. The black triangles represent the data by the HARP [12] experiment.

## 9 Summary

Introducing run time configurable model parameters in Geant4 and developing the configuration interface facilitates variation of the final state content of e.g., hadronic interactions in the detector or beamline simulation. Subsequently it allows to fit simulated distributions to the experimental data, and to extract optimal model parameters as well as the associated correlations and uncertainties. We have initially studied the case with regards to such Geant4 hadronic models as PreCompound, Bertini-like cascade, and FTF. We are currently focusing on a more detailed work on the FTF model parameters, and we have demonstrated that some of its parameters can be optimized through the fitting techniques to bring the Monte Carlo substantially closer to the thin target data, as compared to the default Geant4.11.1 simulation. We plan to continue the study of the FTF parameters with the focus on further improving the agreement between the simulated and the experimental results. We also plan to gradually expand the work on other Geant4 hadronic models. We are considering and exploring a possibility to introduce alternative tunes for FTF and other Geant4 hadronic models, for specific study cases. When matured enough and properly tested such tunes can be offered to users.

## Acknowledgments

† Operated by Fermi Research Alliance, LLC under Contract No. DE-AC02-07CH11359 with the United States Department of Energy.